\documentclass[10pt, paper=a4, UKenglish]{article}
\usepackage{graphicx}
\usepackage{amsmath}
\def\Title#1{\begin{center} {\Large #1 } \end{center}}
\def\Author#1{\begin{center}{ \sc #1} \end{center}}
\def\Address#1{\begin{center}{ \it #1} \end{center}}

\newcommand\pubblock{\rightline{\begin{tabular}{l} Proceedings of the CTD 2023\\ \pubnumber\\
         \pubdate  \end{tabular}}}

\newenvironment{Abstract}{\begin{quotation} \begin{center} 
             \large ABSTRACT \end{center}\bigskip 
      \begin{center}\begin{large}}{\end{large}\end{center} \end{quotation}}

\newenvironment{Presented}{\begin{quotation} \begin{center} 
             PRESENTED AT\end{center}\bigskip 
      \begin{center}\begin{large}}{\end{large}\end{center} \end{quotation}}

\def\Acknowledgements{\bigskip  \bigskip \begin{center} \begin{large}
      \bf ACKNOWLEDGEMENTS \end{large}\end{center}}

\def\beq{\begin{equation}}
\def\eeq#1{\label{#1}\end{equation}}
\def\eeqn{\end{equation}}

\def\beqa{\begin{eqnarray}}
\def\eeqa#1{\label{#1}\end{eqnarray}}
\def\eeqan{\end{eqnarray}}

\let\bar=\overbar

\def\Dslash{\not{\hbox{\kern-4pt $D$}}}
\def\dslash{\not{\hbox{\kern-2pt $\del$}}}

\def\msb{{\bar{\ssstyle M \kern -1pt S}}}

\usepackage{color}
\usepackage{lineno}
\usepackage{subfig}
\usepackage{hyperref}
\usepackage{xspace}
\usepackage[numbers,sort&compress]{natbib}
\usepackage{graphicx}
\graphicspath{{figures/}}

\newcommand{\model}{\textit{TrackingBERT}\xspace}

\newcommand\pubnumber{PROC-CTD2023-33}

\newcommand\pubdate{\today}

\newcommand{\conference}{Connecting the Dots Workshop (CTD 2023)\\
October 10-13, 2023}

\usepackage{fancyhdr}
\definecolor{mygrey}{RGB}{105,105,105}
\begin{document}

% uncomment the following line for adding line numbers
% \linenumbers

% large size for the first page
\large
\begin{titlepage}
\pubblock

%% Change the title, name, abstract
%% Title 
\vfill
\Title{A Language Model for Particle Tracking}
\vfill

%  if you need to add the support use this, fill the \support definition above. 
%  \Author{FIRSTNAME LASTNAME \support}
\Author{Andris Huang~$^1$, Yash Melkani~$^1$, Paolo Calafiura~$^2$, Alina~Lazar~$^3$, Daniel Thomas Murnane~$^2$, Minh-Tuan Pham~$^4$, Xiangyang Ju~$^2$}

\Address{
$^1$ University of California, Berkeley, CA 94720, USA; \\ 
$^2$ Lawrence Berkeley National Laboratory, Berkeley, CA 94720, USA; \\ 
$^3$ Youngstown State University, Youngstown, OH 44555, USA; \\ 
$^4$ University of Wisconsin-Madison, Madison, WI 53706, USA.
}
\vfill

\begin{Abstract}
Particle tracking is crucial for almost all physics analysis programs at the Large Hadron Collider. Deep learning models are pervasively used in particle tracking related tasks. However, the current practice is to design and train one deep learning model for one task with supervised learning techniques. The trained models work well for tasks they are trained on but show no or little generalization capabilities. We propose to unify these models with a language model. In this paper, we present a tokenized detector representation that allows us to train a BERT model for particle tracking. The trained BERT model, namely \model, offers latent detector module embedding that can be used for other tasks. This work represents the first step towards developing a foundational model for particle detector understanding. 

\end{Abstract}

\vfill

% DO NOT CHANGE!!!
\begin{Presented}
\conference
\end{Presented}
\vfill
\end{titlepage}
\def\thefootnote{\fnsymbol{footnote}}
\setcounter{footnote}{0}
%

% normal size for the rest
\normalsize 

%% Your paper should be entered below. 

\section{Introduction}
\label{sec:intro}

Particle tracking plays a vital role in almost all physics programs at the Large Hadron Collider (LHC). It is related to tasks such as lepton reconstruction, jet flavor tagging, primary and displaced vertices reconstruction, and pileup removal for jets and missing energies. Deep learning models are increasingly prevalent in these domains. 
For example, heterogeneous Graph Neural Networks (GNNs) are employed to distinguish hadronically decayed $tau$ leptons from jets~\cite{Huang:2023ssr} with tracks and energy clusters as inputs. 
Sparse transformers is utlized to model secondary proton-proton collisions (commonly referred to as pileup interactions) at the LHC~\cite{Maier:2021ymx}. 
Notably, Ref~\cite{Qu:2022mxj} trained a Transformer~\cite{transformer}-based model on 100 M jet objects evenly distributed in 10 jet classes to classify jets into the ten classes and achieved excellent separation power. 
Ref~\cite{Qu:2022mxj} also demonstrated that fine-tuning pre-trained model can achieve substantially better performance than training the same model from scratch. However, these models are predominantly trained to address specific individual problems or closely related problem sets, primarily using supervised learning techniques.

Large language models (LLMs), such as BERT~\cite{bert}, GPT~\cite{gpt3}, and Llama~\cite{llama:code}, which demonstrated remarkable capabilities in solving various complex tasks, are trained with self-supervised learning. For example, the BERT model, namely ``pre-training of deep directional transformer for language understanding'', is trained for two surrogate tasks: predicting randomly masked words and determining if sentence B logically follows sentence A. This approach enables BERT to achieve state-of-the-art performance in a wide range of sentence-level and token-level tasks. Similarly, GPT models (Generative Pre-trained Transformers) like GPT-2~\cite{gpt2} and GPT-3~\cite{gpt3} use the same Transformer architecture as BERT but differ in their training approach. Unlike BERT's bidirectional context, GPT models are trained using an autoregressive approach, where they predict the next word based solely on the preceding words. This fundamental difference in training methodologies underpins the distinct capabilities of these models in language understanding and generation.

The efficacy of LLMs is significantly influenced by the scaling law~\cite{scaling}. This law states that validation loss is anti-correlated with training data volume, model complexity (i.e., the number of trainable parameters), and computational resources. Although similar trends are observed in applications of Transformers for jet class identification in Ref~\cite{Qu:2022mxj}, the universality of the scaling law across diverse scientific domains remains an area for further exploration.

Particle reconstruction tasks are interconnected. For example a muon object may deposit energy inside a jet. Missing transverse energy can only be obtained once all physics objects are reconstructed. Treating these tasks as distinct entities can lead to a loss of global information, potentially diminishing the overall performance. Algorithms based on physics principles, such as particle flow algorithms~\cite{pflow:atlas, pflow:cms}, endeavored to address this by reconstructing all particles simultaneously. More recently, research efforts~\cite{DiBello:2020bas,Pata:2021oez,Pata:2022wam} aim to replicate this holistic approach using machine learning models. These models are trained on extensive datasets featuring low-level detector information in a supervised learning framework.

In light of the advancements and methodologies employed by LLMs, we propose a novel approach: the development of a generalist intermediary model. This model would be designed to provide learned detector encodings, effectively serving a range of particle reconstruction problems. By harnessing the capabilities demonstrated by LLMs, this model aims to integrate and process the intricacies of particle tracking in a more unified and efficient way.

\section{TrackML detector}
\label{sec:tracking}
Our work uses the TrackML data~\cite{trackml:kaggle, trackml:coda}, which uses top quark pair production from proton-proton collisions as a representative process for track reconstruction at the LHC.  In order to emulate a realistic occupancy for the high-luminosity LHC, a Poisson random number (with mean 200) of minimum bias events are overlaid on top of the $t\bar{t}$ collisions.  This leads to an average of about 10,000 particles/event. The hard-scatter and minimum bias events are both simulated using \textsc{Pythia}~\cite{Sjostrand:2006za,Sjostrand:2007gs}.

\begin{figure}
    \centering
    \subfloat[]{\includegraphics[width=0.43\textwidth]{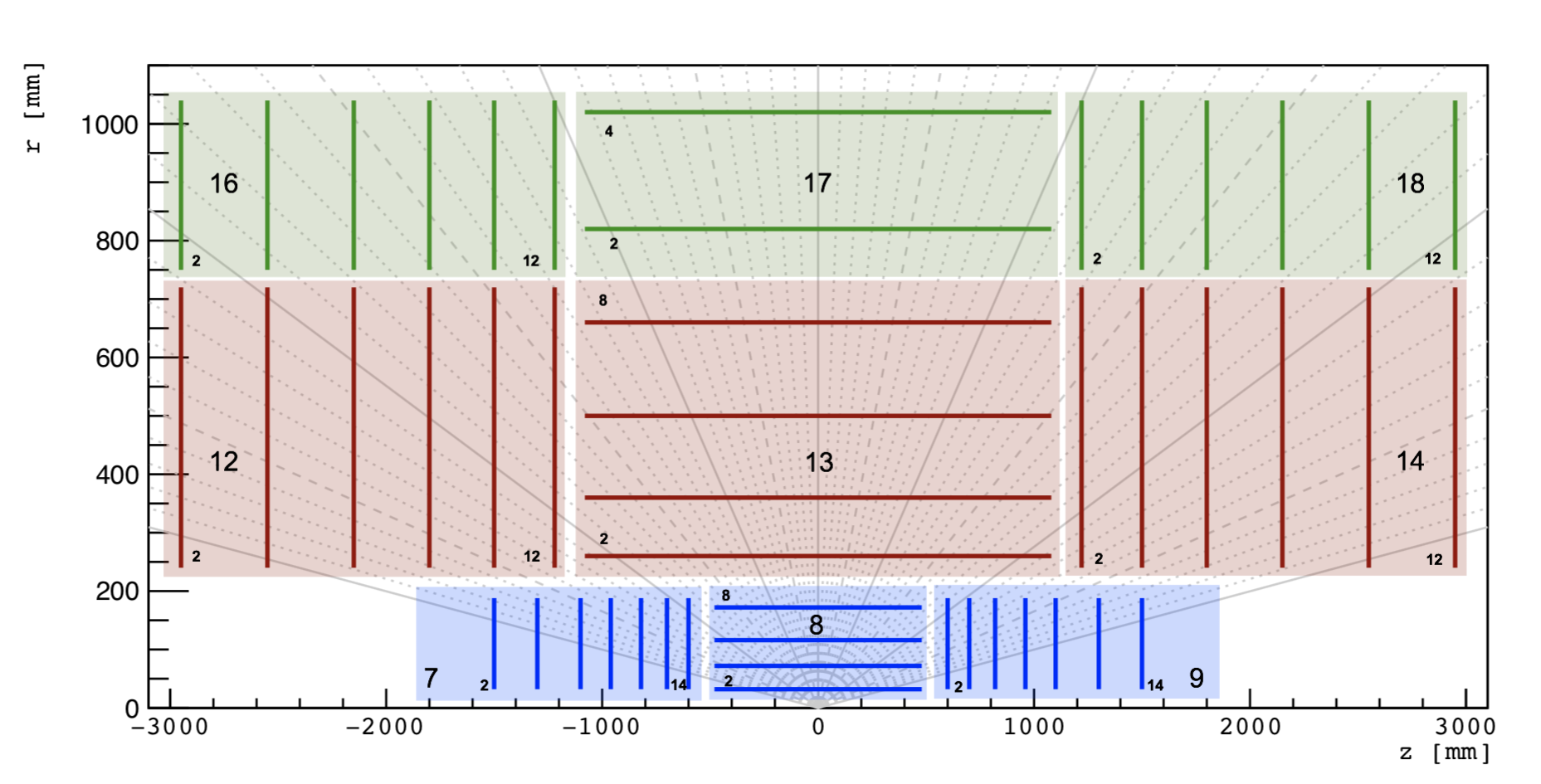}}
    \qquad
    \subfloat[]{\includegraphics[width=0.43\textwidth]{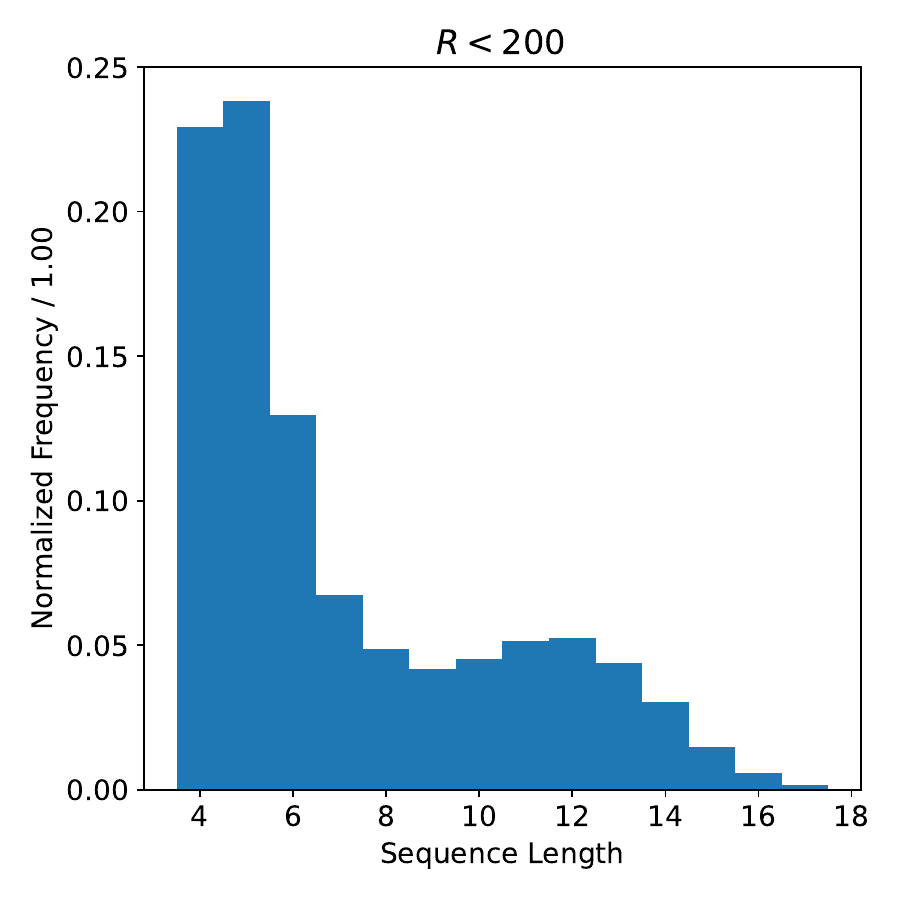}}
    \caption{(a) Layout of the TrackML detector. (b) Number of hits in one track, namely the sequence length in the figure.}
    \label{fig:hits}
\end{figure}

Figure~\ref{fig:hits}(a) shows the layout of the TrackML detector. It is comprised of three main subdetectors: a pixel subdetector (blue), a long strip subdetector (red), and a short strip subdetector (green). The pixel subdetector, which is the focus of our study, consists of three volumes, each containing several layers labeled with even numbers and composed of thousands of modules filled with pixel sensors.

For the following studies, we focus on hits recorded by detector volumn 7, 8, and 9, marked as blue in Fig.~\ref{fig:hits}(a). Figure~\ref{fig:hits} shows the number of hits in one track, namely the sequence length. To further simplify the task, we select all tracks with 4 to 8 hits. This results in a dataset of 4 million tracks, which forms the basis for our training and testing.

\section{Model}
\label{sec:model}

Words are foundational elements in language models like BERT, much as detector readouts are essential in all particle reconstructions. To illustrate this parallel, Table~\ref{tab:analogy} shows an analogy between Natural Language Processing (NLP) and a particle detector.

\begin{table}[!htb]
  \centering
  \begin{tabular}{c|c}
    \hline
    NLP & Particle Detector \\ \hline
    Words & Detector readouts \\
    Vocabulary & All detector readouts \\
    Sentences & Particle trajectories or energy showers \\
    Paragraphs & Collision Events \\
    Sections & Events from the same physics Process \\
    \hline
  \end{tabular}
  \caption{Analogy between Nature Language Processing and a generic detector. \label{tab:analogy}}
\end{table}
Take silicon-based particle tracking detector as an example. Each silicon sensor can be treated as a word, and all silicon sensors in the detector form a vocabulary. Particle trajectories, which are sequences of detector hits, resemble sentences, and a collision event is analogous to a paragraph. 

In language processing, words undergo \textit{tokenization}, a critical step for efficiency in learning~\cite{Toraman_2023} and handling out-of-vocabulary words. The open-source Large Language Model, Llama 2~\cite{llama2}, has a vocabulary of 32k tokens, whereas there are billions of silicon sensors in a detector. The application of tokenization to detector readouts in High Energy Physics (HEP) is an area for further exploration. In our study, we tokenize detector modules with their unique identifiers and use these tokens as inputs to a language model.

Our approach involves training a BERT model with particle tracking data, dubbed as \model. It takes as inputs two tracks randomly sampled from tracks in the training data. Its core component is the Transformer model~\cite{transformer} whose hyperparameters include the number of attention heads $n_h$, the embedding dimensions $d_\text{emb}$, feed forward multi-layer perceptron (MLP) size $l$. After tuning these Transformer hyperparameters, \model achieves good performance with the following hyperparameters: $n_h = 8$, $d_\text{emb} = 64$, $l= 512$, rendering about 1 million trainable parameters. The model is trained on two surrogate tasks: predicting the masked detector modules in a track (MDM) and determining if the second track has higher momentum than the first for 200 epochs (NTP). For the MDM task, a fraction of modules are masked and the \model output is fed to a simple classification layer to predict the missing module. We incrementally increased the mask rate from 15\% to 30\% and then to 50\% during the training. For the NTP task, the \model output is fed to a different classification layer to classify if track A has higher momentum than track B. The model's parameters are updated by the Adam optimizer~\cite{adam} supplemented with weight decay regularization~\cite{adamW}.

\section{Model Performance Evaluation}
\label{sec:res}

The model performance can be evaluated primarily by its ability to predict missing detector hits in a track, aligning with its training objectives. We assess this by masking one module in a track and utilizing \model to predict its location. The prediction's accuracy is measured by the distance between the predicted and actual module locations, determined by their central positions in global coordinates.

\begin{figure}[htb]
  \centering
  \includegraphics[width=0.9\textwidth]{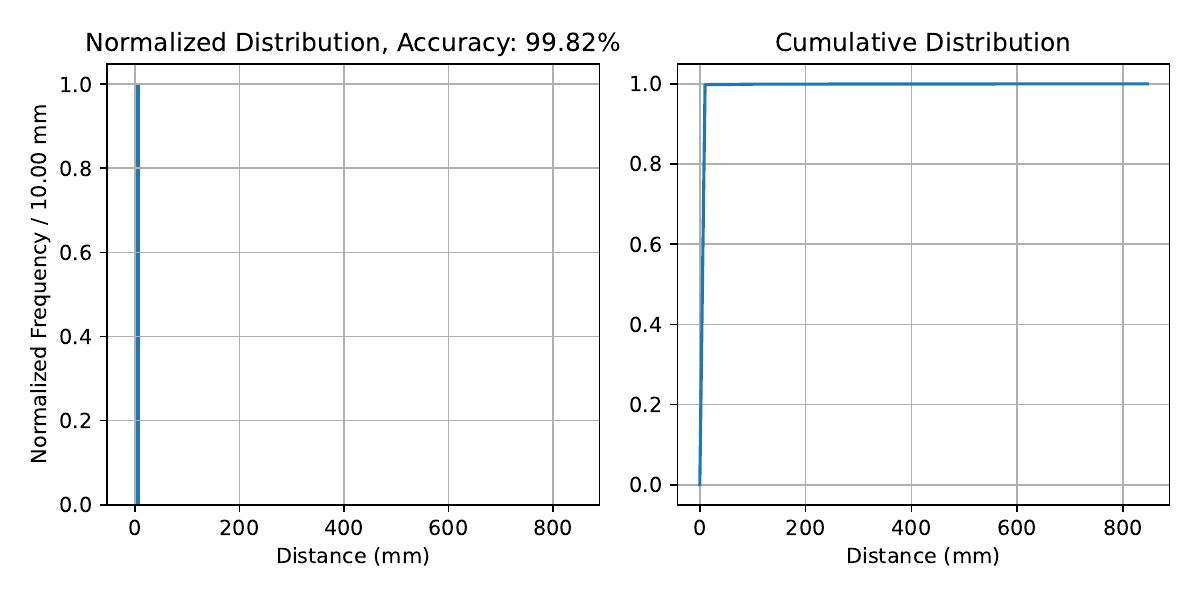}
  \caption{Distances between model predicted module location and the true module location. \label{fig:distance}}
\end{figure}

As showed in Figure~\ref{fig:distance}, the distance distributions for 5000 masked tracks (randomly selected from the testing data) show that the \model precisely predicts the missing module in 99.8\% of cases, achieving 100\% accuracy within a 20 mm range.

\begin{figure}[htb]
  \centering
  \includegraphics[width=0.6\textwidth]{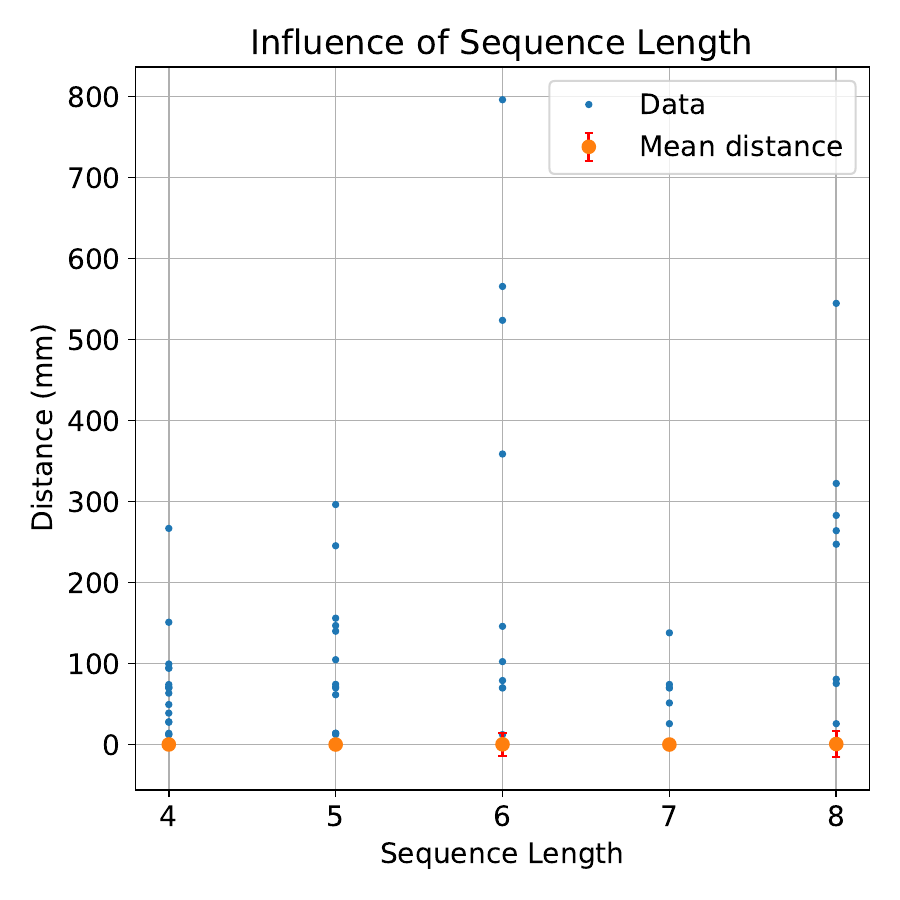}
  \caption{Model performance in predicting masked modules for different track length. \label{fig:dis_vs_seq}}
\end{figure}

Further analysis, shown in Fig.~\ref{fig:dis_vs_seq}, indicates no significant dependency of the model's performance on track length. The model also maintains similar accuracy when predicting masked modules located in different positions in a track.

%%%%%%%%%%%%%%%%%%%%%%%%%%%%%%%%%%%%%%%%%%%%%%%%%%%%%%%%%%%%%%%%%%%%%%%%%%%

\section{Conclusions and Outlooks}
We introduced a novel approach to tokenize detector readout data in High Energy Physics, treating particles as sequences of detector-element tokens. This methodology enabled the direct application of language models in the field. Our \textit{TrackingBERT} model, trained through self-supervised learning, demonstrates high accuracy in predicting masked detector modules. Consistent with the scaling law~\cite{scaling}, we observed enhanced performance with larger datasets and models.

Currently, \model is trained on a simplified dataset. Future work will involve utilizing data from the OpenData Detector~\cite{odd, acts:code} for a more realistic simulation and expanding the training to include full detector modules, such as Strip detectors.

We aim to extract learned detector representations from \model to assess their utility in other tasks, like metric-learning based graph construction in the ExaTrkX pipeline~\cite{ExaTrkX:2021abe}. Additionally, exploring the impact of various masking schemes on Transformer capabilities will be a key focus. For example, masking central modules in tracks teaches the model interpolation, while masking all Strip hits guides it towards extrapolation. While BERT offers flexibility in testing different masking schemes, experimenting with GPT models remains an intriguing prospect.

%%%%%%%%%%%%%%%%%%%%%%%%%%%%%%%%%%%%%%%%%%%%%%%%%%%%%%%%%%%%%%%%%%%%%%%%%%%  

%%  if necessary
\Acknowledgements
This research was supported in part by the U.S. Department of Energy’s Office of Science, Office of High Energy Physics, under Contracts No. DE-AC02-05CH11231 (CompHEP Exa.TrkX) and No. KA2102021235 (US ATLAS Operations), and by the Exascale Computing Project (17-SC-20-SC), a joint project of DOE’s Office of Science and the National Nuclear Security Administration. This research used resources from the National Energy Research Scientific Computing Center (NERSC), a U.S. Department of Energy Office of Science User Facility located at Lawrence Berkeley National Laboratory, operated under Contract No. DE-AC02-05CH11231.

%%%%%%%%%%%%%%%%%%%%%%%%%%%%%%%%%%%%%%%%%%%%%%%%%%%%%%%%%%%%%%%%%%%%%%%%%%%
\bibliography{eprint}
\bibliographystyle{JHEP}
%%%%%%%%%%%%%%%%%%%%%%%%%%%%%%%%%%%%%%%%%%%%%%%%%%%%%%%%%%%%%%%%%%%%%%%%%%%

\end{document}